\documentclass{article}
\usepackage{csquotes}  
\usepackage{arxiv}
\usepackage[pdftex]{graphicx}
\usepackage{amsmath}
\usepackage{placeins}
\usepackage{enumerate}
\usepackage{float}
\usepackage{mcode}   
\usepackage{array} % and/or
\usepackage{longtable} % and/or
\usepackage{colortab} % or
\usepackage{colortbl}
\usepackage{arydshln}
\usepackage{amsthm}
\usepackage{hyperref}
\usepackage{import}
\usepackage{tikz}
\usepackage{bm}
\usepackage{enumitem}
\usepackage[colorinlistoftodos]{todonotes}
\usepackage{algorithm}
\usepackage{algpseudocode}
\usetikzlibrary{matrix,decorations.pathreplacing, calc, positioning}
\newtheorem{theorem}{Theorem}

\begin{document}
\title{Discrete-time Queueing Model of Age of Information with Multiple Information Sources}
\author{
    Nail~Akar\\
	Electrical and Electronics Engineering Dept.\\
	Bilkent University \\
	Bilkent 06800, Ankara, Turkey \\
	\texttt{akar@ee.bilkent.edu.tr} \\
	\And	
    Ozancan~Doğan\\  
    Electrical and Electronics Engineering Dept.\\
	Bilkent University, \\
Bilkent 06800, Ankara, Turkey \\
    \texttt{ozancan@ee.bilkent.edu.tr} \\     
}
	\footnote{Mr. Dogan is supported in part by the {\em 5G and Beyond} scholarship granted by the Information and Communication Technologies Authority (ICTA) of Turkey and Vodafone Turkey. }
\maketitle

\begin{abstract}
	Information freshness in IoT-based status update systems has recently been studied through the Age of Information (AoI) and Peak AoI (PAoI) performance metrics. In this paper, we study a discrete-time server arising in multi-source IoT systems which accepts incoming information packets from multiple information sources so as to be forwarded to a remote monitor for status update purposes. Under the assumption of Bernoulli information packet arrivals and a common geometric service time distribution across all the sources, we numerically obtain the exact per-source distributions of AoI and PAoI in matrix-geometric form for three different queueing disciplines: i)  Non-Preemptive Bufferless (NPB) ii)  Preemptive Bufferless (PB) iii) Non-Preemptive Single Buffer with Replacement (NPSBR).
	The proposed numerical algorithm employs the theory of Discrete-Time Markov Chains (DTMC) of Quasi-Birth-Death (QBD) type and is matrix analytical, i.e, the algorithm is based on numerically stable and efficient vector-matrix operations.
	Numerical examples are provided to validate the accuracy and effectiveness of the proposed queueing model.
	We also present a numerical example on the optimum choice of the Bernoulli parameters in a practical IoT system with two sources with diverse AoI requirements.	
\end{abstract}

\section{Introduction}
In a networked control and monitoring IoT-based system, it is of utmost importance to deliver timely status updates and thus keep the information fresh, for stable operation.
Performance metrics using the Age of Information (AoI) and Peak AoI (PAoI) processes have first been proposed in \cite{kaul_etal_SMAN11,kaul_etal_infocom12,kaul_etal_ciss12} in 
continuous-time,
for quantitative assessment of information freshness in status update systems. Since then, there has been a surge of interest in AoI research in terms of development of queueing models \cite{pappas_etal_icc15,kosta_etal,costa_etal_TIT16,chen_huang_isit16,yates_kaul_tit19,inoue_etal_tit19} or AoI optimization  \cite{sun_etal_tit17,huang_modiano,arafa_ulukus_asilomar17,hsu_etal_isit17,he_etal_TIT18}. 
In the discrete-time setting, for each information source, there is an underlying stochastic process which is randomly sampled, and the sampled values are transmitted in the form of information packets towards a remote monitor via a packet-based communications network.  
Subsequently, for each information source, there is an individual stochastic process, called the AoI process (or sequence), which is maintained at the monitor that keeps track of the time elapsed since the generation of the last successfully received update packet. Therefore, this per-source AoI process is a cyclic process that increases in time with unit steps until the end of a cycle at which a packet reception occurs and subsequently the AoI process experiences an abrupt downward jump. 
On the other hand, the PAoI process is obtained by sampling the AoI sequence at the embedded epochs just before the downward jumps during each cycle of the AoI process \cite{costa_peak}.
\begin{figure}[tb]
	\centering
	\includegraphics[width=0.9\linewidth]{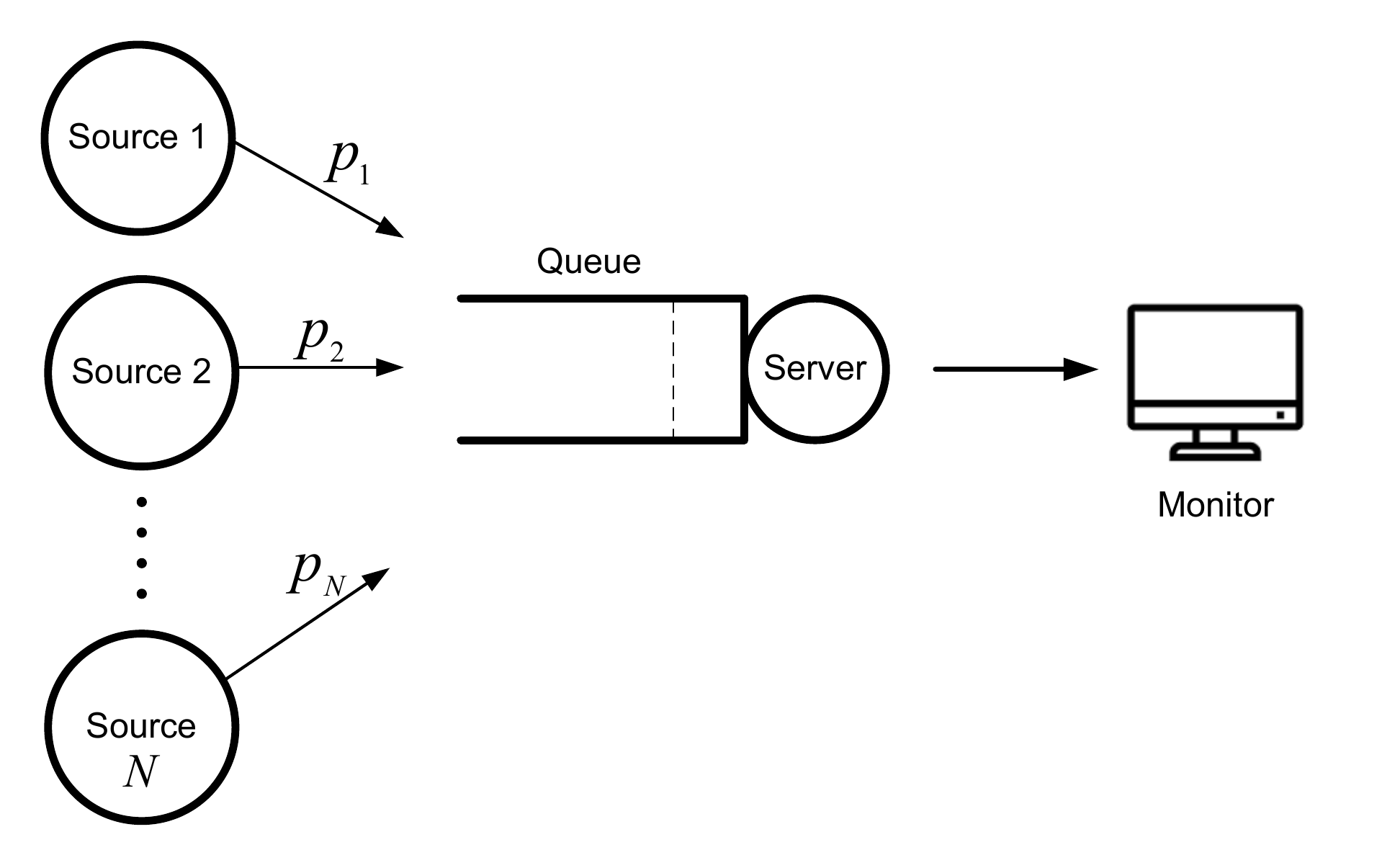}
	\caption{$N$ information sources sending status update messages through a server towards a remote monitor.} 
	\label{fig:multisource}
\end{figure}
The focus of this paper is on an  IoT-based information update system operating in discrete-time in Fig.~\ref{fig:multisource} comprising $N$ independent information sources each equipped with a sensor, a queue-server combination local to the sources, and a remote monitor. In this framework, source-$n$, $n=1,\ldots,N,$ generates information packets according to a Bernoulli process with parameter $p_n$ containing sensed data and a time stamp. The server is in charge of sending the information packets to the monitor via a communications network which introduces random delays, i.e., service time of packets, and the monitor immediately sends back positive acknowledgments to the server. 
We assume in this paper that the service times of all users are geometrically distributed with the same parameter $q$. 
Particularly, we study the following three queueing disciplines employed at the server:
\begin{itemize}
	\item Non-Preemptive Bufferless (NPB) system for which one of the generated information packets at each time instant is selected  uniformly at random, or randomly in short, to be placed in service when the server is idle. Otherwise, all packets are discarded.
	\item Preemptive Bufferless (PB) system in which one of the generated information packets at each time instant (selected randomly) is always placed in service while possibly preempting the one in service. 
	\item Non-Preemptive Single Buffer with Replacement (NPSBR) system for which we have a buffer holding one information packet only and one of the generated information packets 
	at each time instant is randomly selected to replace the one in the waiting room. If there are no packets in the buffer, this information packet is placed in service.
\end{itemize}

The main contribution of this paper is that, based on the theory of QBDs, we propose a novel analytical modeling technique to derive the exact distributions of per-source AoI and PAoI in discrete-time, for NPB, PB, and NPSBR. Although distributional results exist in the literature for continuous-time, results for discrete-time are less mature except for ones that provide average AoI and PAoI values in less general settings. 

The organization of the paper is as follows. In Section~2, related work is presented. Section~3 presents QBDs. Section~4 addresses the proposed queueing model for the bufferless queueing disciplines NPB and PB as well as the single-buffer system NPSBR. Numerical examples are provided to validate the accuracy and effectiveness of the proposed queueing model in Section~5.
Finally, we conclude in Section~6.

\section{Related Work on Queueing Models for AoI and PAoI}
The AoI concept was first introduced in \cite{kaul_etal_infocom12} as a single-source, single-server $M/M/1$ queueing model. 
The case of multiple sources is then investigated in \cite{yates_kaul_ISIT12} in the same setting. Variations of this single-server queueing model in Fig.~\ref{fig:multisource} for AoI has caught the attention of researchers and practitioners for modeling IoT-based status update systems in the literature \cite{kaul_etal_infocom12,kosta_etal_survey}. 
The majority of the existing queueing models on AoI or PAoI are for continuous-time operation. 

Let us first briefly overview the literature on single-source continuous-time models.
In \cite{kaul_etal_infocom12}, the mean AoI is obtained for the single-source $M/M/1$, $M/D/1$, and $D/M/1$ queues with infinite buffer capacity and FCFS scheduling. Although 
current packet-switched communication networks such as the Internet employ large buffers and FCFS scheduling at the routers, their use is shown to have adverse effects on AoI performance figures in moderate to high load regimes \cite{kaul_etal_infocom12}.
The distributions of AoI and PAoI are therefore studied in \cite{costa_etal_TIT16} for the case of small buffers including the conventional $M/M/1/1$ and $M/M/1/2$ queues, as well as the non-preemptive LCFS $M/M/1/2^{\ast}$ queue. 
%The mean AoI and PAoI figures in the pre-emptive LCFS $M/G/1/1$ queueing system is studied in \cite{najm_nasser_isit16}
%where a new arrival preempts the packet in service and the service time distribution is assumed to follow a more general gamma distribution.
Exact expressions for the stationary distributions of AoI and PAoI for a very wide class of single-source information update systems are given in \cite{inoue_etal_tit19}.
The reference \cite{akar_etal_tcom20} obtains the exact distributions of AoI and PAoI in a bufferless $PH/PH/1/1$ system with probabilistic preemption as well as a single-buffer $M/PH/1/2$ system that allows probabilistic replacement of the waiting packet by a newer packet arrival.

Next, we give an overview of the existing literature on multi-source status update systems in continuous-time.  Mean PAoI expressions for $M/G/1$ and $M/G/1/1$ systems with heterogeneous service time requirements are presented in \cite{huang_modiano}. The reference \cite{yates_kaul_tit19} investigates the multi-source $M/M/1$ model with FCFS as well as two disciplines: preemptive bufferless and nonpreemptive single buffer with replacement using the theory of stochastic hybrid systems (SHS) and obtain exact expressions for the mean AoI.
A preemptive $M/G/1/1$ queue is considered in \cite{najm2018status} with a common service time for all sources in which expressions for the mean AoI and PAoI are derived.
The authors of \cite{farazi_etal_Asilomar19} allow self-preemption in which case mean AoI expressions are derived for each source using SHS-based techniques whereas the reference \cite{moltafet2020average} considers a two-source $M/M/1/2$ queueing system in which a packet waiting in the queue can be replaced only by a newly-arriving packet from the same source, again using SHS. A non-preemptive $M/M/1/m$ with common service times across sources is again studied by SHS in \cite{kaul2020timely} and mean AoI expressions are derived. A more general hyper-exponential  service time distribution for each class is considered in \cite{yates_etal_isit19} for an $M/H_2/1/1$ non-preemptive bufferless queue to obtain an expression for the mean AoI per class.    

The existing research on discrete-time AoI queueing models is less mature in comparison with continuous-time models. A discrete-time queueing model with Bernoulli arrivals and geometric service times, using FCFS and non-preemptive LCFS scheduling is presented in \cite{kosta_etal_isit19} with expressions for the mean AoI and PAoI values.
The authors of \cite{tripathi_arxiv} study the FCFS-type $Ber/G/1$ queue and derive explicit expressions for average AoI and PAoI and also mean AoI expressions for the discrete-time LCFS queue. 
For the single-source case, \cite{kosta2020nonlinear} considers the FCFS $Geom/Geom/1$ queue and obtains closed-form expressions of the generating functions and the stationary distributions of the AoI and the PAoI and provide a methodology for analyzing more general non-linear age functions.

\section{Discrete-Time Markov Chains (DTMC) of Quasi-Birth-Death (QBD) Type}
We first present notation. Uppercase bold letters are used to denote real-valued matrices. Lowercase bold (plain) letters or symbols are used to denote real-valued vectors (scalars).
%The $(i,j)^{\text{th}}$ entry of $\bm{A}$ is denoted by ${\bm{A}(i,j)}$ and the $j^{\text{th}}$ entry 
%of a row or column vector $\bm{\alpha}$ is $\bm{\alpha}(j)$. 
The notations $\bm{0}_{k \times \ell} $, ${\bm I_m}$, and ${\bm 1_n}$ denote a matrix of zeros of size $k \times l$, an identity matrix of size $m$, and a column matrix of
ones of size $n$, respectively. When used without a subscript, we leave it to the reader to infer the size information, from the context.  
%The notation $\left[\bm{\alpha},\bm{\beta}  \right]$ is used for the concatenation of the two row vectors $\bm{\alpha}$ and $\bm{\beta}$. 
%A quasi-upper triangular matrix $\bm{A}$  is upper triangular with the exception that it may have either 1-by-1 or 2-by-2 blocks on its diagonal corresponding to real and complex eigenvalues of the matrix $A$, respectively. 
%A square matrix is said to be stable (anti-stable) if each of its eigenvalues has negative (non-negative) real parts. 
The function $u_k$ stands for the discrete-time unit step function, i.e., $u_k=1,k=0,1,\ldots$ and is zero otherwise. The function $\delta_k$ stands for the discrete-time unit impulse function, i.e., $\delta_k=1$ for $k=0$ and is zero, otherwise.

A discrete random variable $X$ is said to possess a matrix geometric (MG) distribution, i.e., $X \sim MG(\bm{c},\bm{A},\bm{b},d)$,
if its probability mass function (pmf), denoted by $p_X(\ell)$, is of the form  \cite{akar_sm15}:
\begin{equation}
p_X(\ell ) = \Pr \{ X=\ell \}= \left\{ \begin{array}{ll}
\bm{c} {\bm A^{\ell-1}}  \bm{b},  & \ell \geq 1, \\
d, & \ell=0.
\end{array} \right.
\label{mg}
\end{equation}
The probability generating function (pgf) of $X$, denoted by $p_X^{\ast}(z)$, is then of the form
\begin{equation}
p_X^{\ast}(z) = \sum_{\ell=0}^{\infty} p_X(\ell) z^{\ell} = \bm{c} (z^{-1}\bm{I} - \bm{A})^{-1} \bm{b} + d. \label{pgfmg}
\end{equation}
The factorial moments of MG distributions 
can be found by differentiating 
\eqref{pgfmg} successively with respect to $z$ and substituting $z=1$. Consequently, the $i^{\text{th}}$ factorial moment
of an MG-distributed random variable $X$ can be found through the following expression (see \cite{akar_sm15}): 
\begin{equation}
E[X(X-1) \cdots (X-i+1)] = i! \bm{c} (\bm{I} - \bm{A})^{-i-1} \bm{A}^{i-1} \bm{b}. \label{moments}
\end{equation}

An infinite QBD-type DTMC \begin{equation}
X_k = (L_k, P_k) \sim QBD(\bm{B_0},\bm{B_1},\bm{A_0},\bm{A_1},\bm{A_2}), \ k \geq 0, \label{qbd}
\end{equation}
is a two-dimensional chain with 
$ (L_k, P_k) \in \{ (i,j): \: 0 \leq i < \infty, \:1 \leq j \leq m \}$, where $L_k$ represents the level sequence of the QBD, $P_k$ stands for the phase sequence, and $X_k$ has an irreducible probability transition matrix $\bm{P}$
of the following canonical block tridiagonal form:
\begin{align}
\bm{P} & = \begin{pmatrix}
\bm{B_0} & \bm{A_0} &  & \\
\bm{B_1} & \bm{A_1} & \bm{A_0} &  \\
& \bm{A_2} & \bm{A_1} & \ddots   \\
&  & \bm{A_2} & \ddots  \\
& & & \ddots
\end{pmatrix},
\label{Pinf}
\end{align}
with $\bm{B_0}$, $\bm{B_1}$, $\bm{A_0}$, $\bm{A_1}$, $\bm{A_2}$, being
$m \times m$ non-negative matrices. Finite QBD chains where the level of the chain stays within a bound is outside the scope of this paper.
The stationary probability vector
$\bm{\pi} = \left[\bm{\pi_0}, \bm{\pi_1}, \ldots \right]$ 
where $\bm{\pi_k}$, of size $1 \times m$, is the solution vector for
level $k$, 
is the unique solution to the following equations
\begin{equation}
\bm{\pi} = \bm{\pi} \bm{P}, \; \sum_{k=0}^{\infty} \bm{\pi_k}  \bm{1}= 1.
\label{piP}
\end{equation}
The stationary solution (when it exists) has a matrix-geometric form~\cite{N81},
i.e.,
\begin{equation}
\bm{\pi_k} = \bm{\pi_0} \bm{R}^k u_k, \label{mg}
\end{equation}
where the matrix-geometric rate matrix $\bm{R}$ (all its eigenvalues being inside the unit circle) is the unique minimal
non-negative solution of the following quadratic matrix equation:
\begin{equation}
\bm{R} = \bm{A_0} + \bm{R A_1 + R^2 A_2}. \label{R1}
\end{equation}
Once $\bm{R}$ is computed, the boundary vector $\bm{\pi_0}$ in (\ref{mg})
can be found  from the following linear matrix equation \cite{N81}:
\begin{equation}
\bm{\pi_0} = \bm{\pi_0} \left( \bm{B_0} + \bm{R B_1} \right) , \;\bm{\pi_0} (\bm{I} - \bm{R})^{-1} \bm{1} = 1. \label{pi0}
\end{equation}
Several algorithms with computational complexity $\mathcal{O}(m^3)$ are known in the literature for obtaining the rate matrix $\bm{R}$ including
the logarithmic reduction procedure~\cite{LR93}, or
the invariant subspace algorithm using the ordered Schur decomposition~\cite{akar_etal_sm00}, both algorithms being computationally efficient and stable. 
In the current paper's numerical examples, we use the latter algorithm.

Let $L$ be the steady-state random variable associated with the level sequence $L_k$ of the QBD with the following expression for its pmf $p_L(\ell)$:
\begin{equation}
p_L(\ell) = \lim_{k \rightarrow \infty} \Pr \{ L_k = \ell\} =  \bm{\pi_0} \bm{R}^{\ell} \bm{1} u_{\ell}.
\end{equation}
Therefore, $L \sim MG(\bm{\pi_0} \bm{R}, \bm{R}, \bm{1}, \bm{\pi_0} \bm{1})$. Another relevant random variable is the steady-state level sequence but restricted to a particular subset of the entire set of phases. 
For this purpose, we define $L^{\mathcal{S}}$ to be the steady-state random variable associated with the level sequence $L_k$ of the QBD restricted to the particular subset $\mathcal{S} \subset \{1,2,\ldots,m \}$. In this case,
\begin{equation}
p_{L^{\mathcal{S}}}(\ell) = \lim_{k \rightarrow \infty} \Pr \{ L_k = \ell \ |  \ P_k  \in \mathcal{S} \}.
\end{equation}
It is not difficult to show that 
\begin{equation}
L^{\mathcal{S}} \sim MG(\alpha \bm{\pi_0} \bm{R}, \bm{R}, \bm{h},\alpha \bm{\pi_0} \bm{h}), \label{restriction}
\end{equation}
where $\bm{h}_i = 1$ if $i \in \mathcal{S}$ and is zero otherwise, and \[ 
\alpha = \left( \bm{\pi_0} (\bm{I} - \bm{R})^{-1} \bm{h} \right)^{-1} .\]
Therefore, factorial moments (hence also ordinary moments) of $L^{\mathcal{S}}$ can easily be computed from the expression \eqref{moments} for any subset $\mathcal{S}$.

%Both of these methods are shown to have similar performances in terms of
%execution time requirements when the latter employs matrix-sign function
%iterations with quadratic convergence rates~\cite{AS_mg1}.
\section{Discrete-Time Queueing Modeling of AoI and PAoI}
We first describe the AoI and PAoI processes (sequences) for the three queueing disciplines NPB, PB, and NPSBR, accompanied by an illustrative example. 
For this purpose, a successful information packet is first defined as one which is received successfully by the monitor; others those that could not start service or that were preempted while in service (this latter situation being specific to PB) are called unsuccessful packets. 
Let $t^{(n)}_j$ denote the arrival instant of the $j^{\text{th}}, j \geq 1$ successful source-$n$ information packet arriving at the server and let $\delta^{(n)}_j,j \geq 1$ denote the reception instant at the monitor of the $j^{\text{th}}$ successful source-$n$ information packet.
We denote by $\Delta^{(n)}_k, k=0,1,\ldots,$ the discrete-time discrete-valued random sequence representing the AoI for source-$n$ at discrete-time instant $k$ with a given initial condition $\Delta^{(n)}_0$. 
\begin{itemize}
	\item When $k \geq 0$, $\Delta^{(n)}_k$ is incremented by one at each time instant until the first source-$n$ successful packet reception occurring  at instant $k=\delta^{(n)}_1$ at which it is again first incremented to yield the PAoI value $\Phi^{(n)}_1$  where $\Phi^{(n)}_j$ denotes the PAoI process for source-$n$ which is a discrete-time discrete-valued random sequence associated with the AoI just before the reception of the $j^{\text{th}}$ successful source-$n$ information packet. 
	\item Moreover, at instant $k=\delta^{(n)}_1$, just after the packet's reception, $\Delta^{(n)}_k$ is set to $D^{(n)}_1$ where $D^{(n)}_j$ is the time spent in service by the $j^{\text{th}}$ successful source-$n$ information packet i.e., $D^{(n)}_j= \delta^{(n)}_j - t^{(n)}_j, j\geq 1$. 
	\item  Following this packet reception,  $\Delta^{(n)}_k$ is again incremented by one at each time instant until the reception of the second source-$n$ successful packet and the AoI random sequence is obtained by repeating this pattern forever. 
\end{itemize}

Discrete-time Markov modeling is quite different than that of continuous-time since in the latter, only one event can happen at a given time. However, in discrete-time, multiple events can happen at the same time instant. Therefore, the sequence of events happening at each time instant is crucial for modeling. Depending on the particular application or implementation, different event sequences might arise. Although the methodology can be easily be extended to different event sequences, we focus on the particular sequence of events at time instant $k, k=0,1,\ldots,$ which is described in Table~\ref{sequence}.
%\begin{enumerate}
%	\item At the beginning of instant $k$, the AoI values $\Delta^{(n)}_k, 1 \leq n \leq N$ are incremented,
%	\item The service time of the packet, say source-$n$ in service, being over or not is checked, if over then the server is placed into the idle state and the AoI value $\Delta^{(n)}_k$ is updated as $D^{(n)}_j$ if this packet turns out to be the  $j^{\text{th}}$ successful source $n$ packet,
%	\item In NPB, if the server is idle, then one of the newcoming packet arrivals is randomly chosen for starting service. Otherwise, all packet arrivals are discarded. 
%	The situation is the same in PB when the server is idle. Otherwise, one of the packet arrivals is randomly chosen for preempting the packet in service. 
%	In NPSBR, one of the packet arrivals is always first written into the waiting room (possibly replacing the one in the waiting room). Subsequently, the packet in the waiting room starts to receive service if the server is idle. I Otherwise, it will be held at the waiting room until the next time instant.
%\end{enumerate}
\begin{table}[ht]
	\centering
	\caption{The assumed sequence of events at time instant $k$}  
	\begin{tabular}{| m{0.60cm}|m{7cm}|}
		\hline \hline
		\bf{Event} & {\bf Description} \\ \hline
		1 & At the beginning of instant $k$, the AoI values $\Delta^{(n)}_k, 1 \leq n \leq N$ are incremented. \\ \hline 
		2 & The service time of the packet, say source-$n$ in service, being over or not is checked, if over then the server is placed into the idle state and the AoI value $\Delta^{(n)}_k$ is updated as $D^{(n)}_j$ if this packet turns out to be the  $j^{\text{th}}$ successful source $n$ packet. \\ \hline 
		3 & \begin{itemize}[leftmargin=*]
			\item In NPB, if the server is idle, then one of the newcoming packet arrivals is randomly chosen for starting service. Otherwise, all packet arrivals are discarded. 
			\item The situation is the same in PB when the server is idle. Otherwise, one of the packet arrivals is randomly chosen for preempting the packet in service. 
			\item In NPSBR, one of the packet arrivals is always first written into the waiting room (possibly replacing the one in the waiting room). Subsequently, the packet in the waiting room starts to receive service if the server is idle. Otherwise, it will be held at the waiting room until the next time instant.
		\end{itemize} \\ \hline
		\hline	\end{tabular}
	\label{sequence}
\end{table}
As stated above, in our model, we assume that only one of the newcoming arrivals can be picked and processed and it is not possible to place one of the newcoming arrivals in service and another one in the waiting room in the case of NPSBR at the same time instant. We leave such extensions for future research.

Table~\ref{illustrative} provides the sample values of the random sequences $\Delta^{(n)}_k, k \geq 0,$ for $n=1,2$ for a two-source example with zero initial conditions for AoI values for the three queueing disciplines NPB, PB, and NPSBR. For this particular example, we assume that source-$1$ packets  arrive periodically at instants $k=1,5,9,\ldots$, with the first five packets indexed {\em 1a-1e} having service times 5, 2, 7, 3, and 1, respectively, and source-$2$ packets also arrive periodically at instants $k=1,7,13,\ldots$, with the first four packets {\em 2a-2d} having service times 4, 4, 2, and 5, respectively. We assume that the packet generated by source-$1$ (source-$2$) is to be chosen when the two sources generate packets simultaneously at $k=1$ ($k=13$). The following explanations are given for each of the three disciplines to follow the sample evolution of the AoI and PAOI sequences.
\begin{itemize}
	\item For NBP, at $k=1$, packet {\em 1a} is placed in service and $\Delta^{(1)}_k$ is initially incremented by one at each instant until the reception of {\em 1a} at $k=6$ leading to $\Phi^{(1)}_1=6$. 
	In the meantime, packet {\em 1b} is discarded at $k=5$ since the server is busy opon its arrival. Subsequently, $\Delta^{(1)}_k$ is again incremented by one but rising from the value 5 (service time of {\em 1a}) until the reception of {\em 1e} at $k=18$.
	The packets {\em 1b} and {\em 1c} are discarded since they found the server busy and packet {\em 1d} is not picked against packet {\em 2c}.
	Similarly, packet {\em 2a} is not picked against {\em 1a} and packet {\em 2b} is placed in service at $k=4$ and therefore $\Delta^{(2)}_k$ is initially incremented by one at each instant until the reception of {\em 2b} at $k=11$ giving rise to $\Phi^{(2)}_1=11$. Subsequently, $\Delta^{(2)}_k$ is again incremented by one but rising from the value 4 (service time of {\em 2b}) until the reception of {\em 2c} at $k=15$.
	\item For BP, at $k=1$, packet {\em 1a} is placed in service but is preempted by {\em 1b} which is received at $k=7$.
	Thus, $\Delta^{(1)}_k$ is incremented by one until $k=7$ leading to $\Phi^{(1)}_1=7$. Packet {\em 2b} joins service at $k=7$ which is preempted by {\em 1c} which is also eventually preempted by the successful packet {\em 2c} received at $k=15$. Therefore, $\Delta^{(2)}_k$ is incremented by one at each instant until $k=15$ giving rise to $\Phi^{(2)}_1=15$. Following this, packet {\em 1e} joins service at $k=17$ and is received at $k=18$. Thus 
	Subsequently, $\Delta^{(1)}_k$ is again incremented by one but rising from the value 5 (service time of {\em 1a}) until the reception of {\em 1e} at $k=18$.
	Thus, $\Delta^{(1)}_k$ is incremented by one until $k=18$ leading to $\Phi^{(1)}_2=13$ while rising from the value of 2 at $k=7$. 
	\item For NPSBR, at $k=1$, packet {\em 1a} is placed in service and is received at $k=6$ whereas {\em 1b} joins the waiting room at $k=5$ and starts receiving service at $k=6$ which gets to complete at $k=8$.
	Packet {\em 2b} joins the waiting room at $k=7$ and starts to receive service at $k=8$. Thus, the first successful source-$2$ reception occurs at $k=12$ at which packet {\em 1c} starts receiving service which gets to complete at $k=19$. Packet {\em 2c} joins the waiting room at $k=13$ but is replaced with packet {\em 1e} which joins service at $k=17$ and subsequently completes at $k=18$. 
\end{itemize}

\begin{table*}[ht]	
	\centering
	\caption{Sample evolution of the AoI and PAoI sequences for $0 \leq k \leq 20$ for an illustrative example for each of the three queueing disciplines NPB, PB, and NPSBR.}  
	\begin{tabular}{|c|ll|ll|ll|}
		\hline \hline
		& \multicolumn{2}{c|}{\bf{NPB}} & \multicolumn{2}{c|}{{\bf PB}} & \multicolumn{2}{c|}{{\bf NPSBR}
		} \\ \cline{2-7}
		Instant $k$ & $\Delta^{(1)}_k$ &  $\Delta^{(2)}_k$ & $\Delta^{(1)}_k$ &  $\Delta^{(2)}_k$ & $\Delta^{(1)}_k$ &  $\Delta^{(2)}_k$ \\ \hline
		0 & 0 & 0 & 0 & 0 & 0 & 0 \\ \hline
		%	1& 1 & 1 & 1& 1& 1 & 1\\ \hline
		%	2 & 2 & 2 & 2& 2 & 2 & 2 \\ \hline
		%	3 & 3 & 3 & 3 & 3 & 3 & 3  \\ \hline
		%	4 & 4 & 4 & 4 & 4 & 4 & 4  \\ \hline
		$\vdots$ & $\vdots$ & $\vdots$ & $\vdots$ & $\vdots$ & $\vdots$ & $\vdots$  \\ \hline	
		5 & 5 & 5 & 5 & 5 & 5 & 5  \\ \hline 
		6 & 5 ($ 
		\Phi^{(1)}_1  \leftarrow 6$ ) & 6 & 6 & 6 & 5 ($
		\Phi^{(1)}_1  \leftarrow 6)$ & 6 \\ \hline 
		7  & 6  & 7 & 2 ($
		\Phi^{(1)}_1 \leftarrow 7) $ & 7 & 6 & 7\\ \hline 
		8 & 7  & 8  & 3 & 8 & 2 ($
		\Phi^{(1)}_2 \leftarrow 7)$ & 8 \\ \hline 
		9 & 8  & 9 & 4  & 9 & 3 & 9 \\ \hline 
		10 & 9  & 10  & 5 &10  &4 & 10 \\ \hline 
		11 & 10 & 4    $(\Phi^{(2)}_1 \leftarrow 11) $  & 6 & 11 & 5& 11\\ \hline
		12 &  11 & 5 &7 & 12  &6 & 4 ($ \Phi^{(2)}_1 \leftarrow 12)  $ \\ \hline 
		13 & 12 & 6 &8 & 13 &7 &5 \\ \hline  
		14 & 13 & 7 &9 & 14  &8  & 6\\ \hline 
		15 & 14 & 2 $(\Phi^{(2)}_2 \leftarrow 8)$ &10 & 2 ($  \Phi^{(2)}_1 \leftarrow 15) $ & 9  & 7 \\ \hline 
		16 & 15 & 3 & 11& 3& 10& 8\\ \hline
		17 & 16 & 4  &12  & 4  &11 & 9\\ \hline  
		18 & 1 ($ 
		\Phi^{(1)}_2 \leftarrow 17) $ & 5 & 1 ($ 
		\Phi^{(1)}_2 \leftarrow 13) $ &5  &12 & 10\\ \hline 
		19 & 2 & 6 & 2 & 6 & 10 ($
		\Phi^{(1)}_3 \leftarrow 13 $) & 11\\ \hline
		20 & 3  & 7 & 3 & 7  &11 & 12 \\ \hline  
		$\vdots$ & $\vdots$ & $\vdots$ & $\vdots$ & $\vdots$ & $\vdots$ & $\vdots$ \\ \hline \hline	
	\end{tabular}
	\label{illustrative}
\end{table*}

We use the notation $\Delta^{(n)}$, $\Phi^{(n)}$, and $D^{(n)}$, to denote the steady-state random variables associated with the processes $\Delta^{(n)}_k$, $\Phi^{(n)}_j$, and $D^{(n)}_j$, respectively.
For ease of notation, we tag a single source out of all sources, say source-$1$, and drop the superscript while writing the steady-state pmf for the random variables $\Delta = \Delta^{(1)}$, $\Phi = \Phi^{(1)}$, and $D = D^{(1)}$, respectively:
\begin{align}	
p_{\Delta}(\ell ) &= \lim\limits_{k\to \infty }   \Pr \{ \Delta^{(1)}_k =\ell \}, \ \ell \geq 0, \label{pmfaoi} \\
p_{\Phi}(\ell ) &= \lim\limits_{j \to \infty }   \Pr \{ \Phi^{(1)}_j = \ell \}, \ \ell \geq 0, 
\label{pmfpaoi} \\
p_{D}(\ell ) &= \lim\limits_{j \to \infty }   \Pr \{ D^{(1)}_j = \ell \}, \ \ell \geq 0. \label{pmfwait}
\end{align}
We also denote by  $F_{\Delta^{(n)}}(\ell)$, $F_{\Phi^{(n)}}(\ell)$, and $F_{D^{(n)}}(\ell)$, the corresponding cumulative distribution functions (cdf) of the random variables $\Delta^{(n)}$, $\Phi^{(n)}$, and $D^{(n)}$, respectively.
The goal of this paper is to devise a numerical algorithm to write the first two pmfs given in \eqref{pmfaoi} and \eqref{pmfpaoi} or their corresponding cdfs. The pmf given in \eqref{pmfwait} is auxiliary and is to be needed for the former two pmfs for the NPSBR scenario. If the interest is on the two pmfs related to another information source-$n$ where $n \neq 1$, the same procedure can be repeated by renumbering the sources.

Our proposed methodology relies on constructing an infinite Markov chain of QBD type that produces cycles repeating forever in such a way that one cycle begins with the arrival of a successful source-$1$ packet and evolves until the reception of the next successful class-$1$ packet. The exception to this is the PB system where a cycle is to begin with the arrival of a source-$1$ packet which is not necessarily successful. The level is always incremented until the second successful source-$1$ packet reception occurs at which we transition to an auxiliary state where the level is always decremented until the level zero is hit in order to prepare for starting the next cycle. We will show that the steady-state distribution of this properly constructed QBD enables one to find the two pmfs given in \eqref{pmfaoi} and \eqref{pmfpaoi} for each queueing discipline of interest.
Recall that source-$n$ information packet generation is governed by a Bernoulli process with parameter $p_n$ with $\bar{p}_n = 1-p_n$. We also let $p=\sum_{n=1}^N p_n$. The service times of all users are geometrically distributed with the same parameter $q$ and $\bar{q} = 1-q$. The load $\rho$ is defined as $\rho = p/q$. 

\begin{table}[tb]
	\centering
	\caption{Description of the five phases used for NPB.}  
	\begin{tabular}{| m{0.6cm}|m{7cm}|}
		\hline \hline
		\bf{Phase} & {\bf Description} \\ \hline
		1 & The first source-$1$ packet is in service \\ \hline 
		2 & The service of first source-$1$ packet is over and the server is idle \\ \hline 
		3 & The service of first source-$1$ packet is over and the second source-$1$ packet is in service \\ \hline
		4 & The service of first source-$1$ packet is over and a source-$n$ packet is in service where $n \neq 1$ \\ \hline
		5 & The service of the second source-$1$ packet is over and we prepare for the next cycle \\ \hline \hline	\end{tabular}
	\label{5states}
\end{table}
\subsection{Non-Preemptive Bufferless (NPB) Queueing System}
For NPB, we propose an infinite Markov chain $X_k =(L_k,P_k), k=0,1,\ldots$ of QBD type characterized as in \eqref{qbd} with 5 phases. Table~\ref{5states} describes each of the five phases used for the NPB system.
We first define the pgf (probability generating function) of the number of information packet arrivals from all sources other than source 1:
\begin{equation}
\tau(z)=\sum_{j=0}^{N-1} \tau_j z^j = \prod_{n=2}^N (\bar{p}_n + p_n z). \label{pgf}
\end{equation}
Next, we define $\gamma_0= \prod_{n=1}^N \bar{p}_n$ to be the probability of no packet arrivals at a time instant, \[
\gamma_1 = \sum_{j=0}^{N-1} p_1 \frac{\tau_j}{j+1},
\]
to be the probability of a source-$1$ packet to be chosen (uniformly at random) among all packet arrivals, $\gamma_2 = 1 - \gamma_0 - \gamma_1$, to be the probability of a source-$n$ packet ($n \neq 1$) to be chosen among all packet arrivals. The proposed QBD evolves in the form of repetitive cycles each of which begins with the arrival of a source-$1$ packet into an empty system and continues until the reception of the next successful source-$1$ packet. At the beginning of a cycle, the level is zero and we are at phase 1. 
With probability $\bar{q}$, we stay in phase 1 and we transition to phase $i+2$ with probability $q \gamma_i$ while incrementing the level. While in phase 2, we stay at phase 2 until an information packet arrival occurs. Therefore, in phase 2, we transition to phase 3 with probability $\gamma_1$, or to phase 4 with probability $\gamma_2$. The level is still incremented in all these transitions at phase 2.
At phase 3, we transition to phase 5 with probability $q$ when the service time of the next successful source-$1$ packet is over and we stay at phase 3, otherwise. The level is still incremented in phase 3. While in phase 4, we either stay at phase 4, or to phase 3 when the service completes and a source-$1$ packet arrival is chosen at the same instant, or to phase 2 when the service completes but there are no new packet arrivals. Again, the level is incremented in phase 4.
When in phase 5, we stay at phase 5 while decrementing the level until the level zero is hit. When the level is zero, we make a transition to phase 1 with probability one while staying at level zero. This epoch is where a new cycle gets to begin.
With this description, the characterizing matrices of the QBD are as follows:
% $\gamma_{01}=\gamma_0 + \gamma_1 $, $\gamma_{02}=\gamma_0 + \gamma_2$, and $\gamma_{12}=\gamma_1 + \gamma_2$.
\begin{align}
\bm{A_0} & =
\begin{pmatrix}
\bar{q} &q\gamma_0 &q\gamma_1 &q\gamma_2 & 0 \\ 
0& \gamma_0 & \gamma_1&\gamma_2 &0 \\
0& 0& \bar{q}&0 & q \\ 
0& q\gamma_0&q \gamma_1 &  \bar{q} +q \gamma_{2} &0 \\ 
0&0 & 0&0 & 0\\ 
\end{pmatrix},
\label{A0npb}
\end{align}
$\bm{A_1}$ is a matrix of zeros, $\bm{A_2}$
and $\bm{B_1}$ are matrices of zeros except for their $(5,5)^{\text{th}}$ entry which is one, and 
$\bm{B_0}$ is a matrix of zeros except for its $(5,1)^{\text{th}}$ entry which is one.
From the evolution of this QBD, we observe that the values that the level sequence $L_k$ of the proposed QBD takes in phases 2, 3, and 4, in one QBD-cycle, coincide with the sample values of the AoI process in its own cycles, as given in Table~\ref{illustrative}. 
Recall that an AoI cycle starts with the age taking its initial value which is the service time of a successful source-$1$ packet which is then subsequently incremented until the reception of the next successful source-$1$ packet. 
This observation leads us to the following main result for the NPB queueing system.
\begin{theorem}
	\label{thm:npb}
	Let $X_k \sim QBD(\bm{B_0},\bm{B_1},\bm{A_0},\bm{A_1},\bm{A_2})$ having the characterizing matrices as in \eqref{A0npb} with its stationary vector for level $k$ of size $1 \times 5$ denoted by $\bm{\pi_k}$ being in matrix geometric form $
	\bm{\pi_k} = \bm{\pi_0} \bm{R}^k u_k$. Then, the steady-state pmf of the AoI sequence for the NPB system for source-$1$ is the same as that of $L^{\mathcal{S_A}}$ which is the steady-state level of the QBD restricted to the subset $\mathcal{S_A} = \{ 2,3,4\}.$ Moreover, the steady-state pmf of the PAoI sequence for the NPB system for source-$1$ is the same as that of $1+L^{\mathcal{S_P}}$ where  $L^{\mathcal{S_P}}$ is the steady-state level of the QBD restricted to phase 3, i.e., $\mathcal{S_P} = \{ 3 \}.$
	%	\begin{equation}
	%	p_{\Delta}(\ell) = \bm{c_a} \bm{R}^{\ell} \bm{b_a} u_{\ell}, \
	%	p_{\Phi}(\ell ) = \bm{c_p} \bm{R}^{\ell - 1} \bm{b_p} u_{\ell - 1},
	%	\label{thmnpb}
	%	\end{equation}
	%	where \begin{align}
	%	\bm{b_a} & =\begin{bmatrix} 0 & 1 & 1 & 1 & 0 \end{bmatrix}^T, \ \bm{c_a}  = \bm{\pi_0}/ (\bm{\pi_0} (\bm{I} - \bm{R})^{-1} \bm{b_a}), \nonumber  \\ \bm{b_p} &=\begin{bmatrix} 0 & 0 & 1 & 0 & 0 \end{bmatrix} ^T,  \ \bm{c_p} = \bm{\pi_0}/ (\bm{\pi_0}(\bm{I} - \bm{R})^{-1} \bm{b_p}). \label{pmfnpb} 
	%	\end{align}
	%	Moreover, their expected values are given by:
	%	\begin{align}
	%	E[\Delta] & = \bm{c_a} \left( (\bm{I} - \bm{R})^{-2} - (\bm{I} - \bm{R})^{-1}) \right) \bm{b_a}, \nonumber \\
	%	E[\Phi] & = \bm{c_p} \left( (\bm{I} - \bm{R})^{-2} - (\bm{I} - \bm{R})^{-1}) \right) \bm{b_p}. \label{meannpb}
	%	\end{align}
\end{theorem}
\begin{proof}
	The proof is based on sample path arguments.
	From the evolution of this QBD, we observe that the values that the level sequence $L_k$ of the proposed QBD takes in phases 2, 3, and 4, in one QBD-cycle, coincide with the sample values of the AoI process in its own cycles, as given in Table~\ref{illustrative}. 
	%	 We observe that one cycle of the level sequence $L_k$ restricted to the phases 2, 3, and 4, of the QBD, coincides with one cycle of the AoI sequence $X_k$ and vice versa, as given in Table~\ref{illustrative}. 
	Similarly, one added to the values that the level sequence $L_k$ of the proposed QBD takes in phase 3, in one QBD-cycle, coincide with the sample values of the PAoI sequence in one AoI cycle, as given in Table~\ref{illustrative}. 
	%	Similarly, the PAoI sequence coincides with the the level sequence conditioned on phase 3. 
	The pmfs of the AoI and PAoI sequences can then explicitly be written in matrix geometric form as in \eqref{restriction} and their factorial moments can explicitly be written as  in \eqref{moments}.
	%	For the AoI sequence, we need to censor the phases 1 and 5 (or equivalently condition on the phases 2, 3, and 4), and then normalize to yield $\sum_{\ell}p_{\Delta}(\ell)=1$. Similarly, for the PAoI sequence, we need to censor the phases 1, 2, 4, and 5, and then normalize to make $\sum_{\ell}p_{\Phi}(\ell)=1$ but shift right to one unit in the end. 
	%	The reason for the latter operation is that the PAoI sequence is to be obtained by means of adding one to the value of the level process conditioned on phase 3. 
	%	The equations \eqref{thmnpb}-\eqref{pmfnpb} follow from the observation that $\sum_{\ell =0}^{\infty} \bm{R}^{\ell} = (\bm{I} - \bm{R})^{-1}$. On the other hand,  \eqref{meannpb} follows from the observation that $\sum_{\ell =0}^{\infty} \ell \bm{R}^{\ell} = (\bm{I} - \bm{R})^{-2} -(\bm{I} - \bm{R})^{-1}$ for a legitimate rate matrix $\bm{R}$.
\end{proof}

\subsection{Preemptive Bufferless (PB) Queueing System}
For PB, we propose an infinite Markov chain $X_k = (L_k, P_k), k=0,1,\ldots$ of QBD type characterized as in \eqref{qbd} with 5 phases, the description of the first four phases being the same as that of NPB. On the other hand, for PB, in phase 5, either the service of the second source-$1$ packet is over and we prepare for the next cycle, or the first source-$1$ packet is unsuccessful, i.e., preempted during phase 1, and after transitioning to phase 5, we prepare for the arrival of the first source-$1$ packet.
This QBD evolves in the form of repetitive cycles each of which begins with the beginning of a source-$1$ packet's service and continues until the preemption of this packet or reception of the next successful source-$1$ packet. If the source-$1$ packet is unsuccessful during phase 1, i.e., preempted, then the cycle is called an unsuccessful cycle.
Otherwise, the cycle continues until the reception of the second source-$1$ information packet and is called a successful cycle. 
At the beginning of a cycle, the level is zero and we are at phase 1 and the service of the source-$1$ packet is about to start. 
To describe the operation of the QBD, we first need the following definition:
$\gamma_{01}=\gamma_0 + \gamma_1 $, $\gamma_{02}=\gamma_0 + \gamma_2$, and $\gamma_{12}=\gamma_1 + \gamma_2$.
The differences from the NPB system are now listed below in relation to the behavior at phases 1, 3, and 4:
\begin{itemize}
	\item At phase 1, with probability $\bar{q}\gamma_{12}$, the first source-$1$ packet gets to be preempted and we transition to phase 5 whereas the behavior at phase 5 is identical to that of the same phase in NPB.  In this case, the cycle is an unsuccessful cycle. The other transitions to phases 2, 3, and 4, are identical to that of NPB.
	\item In phase 3, if the service of source-$1$ packet successfully completes, we transition to phase 5, which occurs with probability $q$. On the other hand, if this packet is preempted by a source-$n$, $n \neq 1$ packet, we transition to phase 4, which occurs with probability $\bar{q}\gamma_{2}$. Otherwise, we stay at phase 3. 
	\item While in phase 4, a source-$n$ packet, $n \neq 1$, is in service. We transition to phase 3 if a source-$1$ packet is chosen (with probability $\gamma_1$) irrespective of whether the service is over or not. Additionaly, we transition to phase 2 if the service completes and there are no new arrivals, which occurs with probability $q \gamma_0$.
\end{itemize}
%Since unsuccessful cycles only consists of phase 1 and 5 and we censoring these phases from Theorem 1, these cycles are not taken into consideration for the calculation of pmf of AoI and PAoI. 
%%\item If the first source-$1$ packet successfully completes its service with probability $q$, we transition to one of the three phases 2, 3, and 4, with probabilities $q\gamma_0$, $q\gamma_1$, and $q\gamma_2$, respectively. While in phase 2, we wait for an arrival of a packet.  When the chosen packet is a source-$1$ packet, we transition to phase 3, otherwise we transition to phase 4. While in phase 3, if the service of source-$1$ packet successfully completes, we transition to phase 5. On the other hand, if this packet is preempted by a source-$n$ packet, we transition to phase 4. While in phase 4, we either stay at phase 4 or we transition to phase 3 (4) if the packet is preempted by a new class-$1$ (class-$n$), or the service of the source-$n$ packet is completed and a source-$1$ packet is chosen a, or to phase 2 if the service of packet is completed and no packet arrives. For a successful cycle, we transition to phase 5 and we stay at the phase 5 until the level 0 is hit as in the unsuccesful cycle. 
%From the evolution of this QBD, sample values of the AoI coincide with values in phase 2,3 and 4 as in the NPB system. As a difference from NPB system AoI cycle starts with the initial value which is service time of a successful source-1 packet since the preemption renders the cycle as unsuccessful and increases until the reception of the next successful source-1 packet.
The characterizing matrices of the proposed QBD, $X_k$, are thus written as:
\begin{align}
\bm{A_0} & =
\begin{pmatrix}
\bar{q} \gamma_0 &q\gamma_0 &q\gamma_1 &q\gamma_2 & \bar{q}\gamma_{12}\\ 
0& \gamma_0 & \gamma_1&\gamma_2 &0 \\
0& 0& \bar{q}\gamma_{01}&\bar{q}\gamma_{2} & q \\ 
0& q\gamma_0& \gamma_1 & \bar{q}\gamma_{02	} + q \gamma_{2} &0 \\ 
0&0 & 0&0 & 0\\ 
\end{pmatrix},
\label{A0pb}
\end{align}
$\bm{A_1}$ is matrix of zeros, $\bm{A_2}$
and $\bm{B_1}$ are matrices of zeros except for their $(5,5)^{\text{th}}$ entry which is one, and 
$\bm{B_0}$ is a matrix of zeros except for its $(5,1)^{\text{th}}$ entry which is one.
Theorem~1 stated for NPB also applies to PB with the exception that the $\bm{A_0}$ matrix of the QBD for PB is given as in \eqref{A0pb} as opposed to 
$\bm{A_0}$ written as in \eqref{A0pb} for NPB.

\subsection{Non-Preemptive Single Buffer (NPSBR) Queueing System with Replacement}
Before introducing the QBD for NPSBR as the other two queueing systems, we first need to obtain the pmf of the waiting time $p_D(\ell)$ of successful source-$1$ packets in NPSBR. Next, we present the corresponding main result in the following theorem.
\begin{theorem} The pmf of the queue waiting time for successful source-$1$ information packets in NPSBR is a mixture of a probability mass at zero and a geometric distribution, i.e., there exist parameters $a,b, \ 0 \leq a,b \leq 1$ such that the following hold:
	\begin{equation} 
	p_D(\ell) = a \delta_l + (1-a) b (1-b)^{l-1} u_{l-1}. \label{expressionqueuewait}
	\end{equation}
\end{theorem}
\begin{proof}
	Let the random sequence $Y_k \in \{0,1,2\}, k \geq 0$ represent the number of information packets in the NPSBR system (service and waiting room) at instant $k$. It is not difficult to show that the random sequence $Y_k$ is governed by a DTMC with probability transition matrix
	\[
	\bm{Q}=\begin{pmatrix} 
	\gamma_0 & 1-\gamma_0 & 0 \\
	q \gamma_0 & q \gamma_{12} + \bar{q} \gamma_0 &  \bar{q} \gamma_{12} \\
	0 & q & \bar{q} \end{pmatrix}.
	\]
	Let $\bm{x} = \begin{pmatrix}
	{x_0} & {x_1} & {x_2}
	\end{pmatrix}$ be the stationary vector of this DTMC satisfying
	\(
	\bm{x} = \bm{x} \bm{Q}, \; \bm{x}  \bm{1}= 1.
	\)
	Also let $\gamma$ be the probability of a source-$1$ packet to be picked among all other arrivals given that the source-$1$ packet has arrived. Clearly, $\gamma = \gamma_1/ p_1$. Let $p_s$ be the success probability of a source-$1$ packet. An arriving packet joins the server with probability $\gamma (x_0 + {x_1} q + {x_2} q)$ and joins the waiting room with probability $\gamma ({x_1} \bar{q} + {x_2} \bar{q})$. A packet that joins the waiting room is successful iff there are no other arrivals in a geometrically distributed interval with parameter $q$, which occurs with probability $r$. Therefore,
	\[
	p_s = \gamma (x_0 + {x_1} q + {x_2} q) + \gamma r ({x_1} \bar{q} + {x_2} \bar{q}).
	\]
	Consequently, we have the following closed-form expression for $r$:
	\begin{equation*}
	r = \sum_{k=1}^{\infty} \gamma_0^k \bar{q}^{k-1}q
	= \frac{\gamma_0 q}{1-\gamma_0 + \gamma_0 q}. 
	\end{equation*}
	The conditional probability of zero queue wait, denoted by $p_D(0)$, conditioned on a successful source-$1$ packet is $\gamma ({x_0} + {x_1} q + \bm{x_2} q)/p_s$. Similarly, the conditional probability of a queue wait of $\ell$ instants, denoted by $p_D(\ell)$, conditioned on a successful source-$1$ packet is $\gamma ({x_1} \bar{q} + \bm{x_2} \bar{q}) \gamma_0^{\ell} \bar{q}^{\ell -1} q/p_s$. Subsequently, the following choices of the two parameters $a$ and $b$
	\begin{equation}
	a= \gamma ({x_0} + {x_1} q + {x_2} q)/ p_s, \ b=1-\gamma_0 \bar{q},
	\end{equation}
	give rise to the expression \eqref{expressionqueuewait}.
\end{proof}
\begin{table}[t]
	\centering
	\caption{Description of the ten phases used for NPSBR.}  
	\begin{tabular}{| m{0.55cm}|m{6.5cm}|}
		\hline \hline
		\bf{Phase} & {\bf Description} \\ \hline
		1 & The first successful source-$1$ packet is in the waiting room \\ \hline 
		2 & The first successful source-$1$ packet is in service and the waiting room is empty \\ \hline
		3 & The first successful source-$1$ packet is in service and there is a source-$1$ packet in the waiting room \\ \hline 
		4 & The first successful source-$1$ packet is in service and there is a source-$n$ packet in the waiting room with $n \neq 1$ \\ \hline
		5 & The service of first source-$1$ packet is over and we wait for a packet arrival \\ \hline
		6 & The second successful source-$1$ packet is in service \\ \hline
		7 & A source-$n$ ($n \neq 1$) packet is in service and the waiting room is empty \\ \hline 
		8 & A source-$n$ ($n \neq 1$) packet is in service and  and there is a source-$1$ packet in the waiting room \\ \hline 
		9  & A source-$n$ ($n \neq 1$) packet is in service and  and there is a source-$n$ packet ($n \neq 1$) is in the waiting room \\ \hline 
		10 & The service of the second successful source-$1$ packet is over and we prepare for the next cycle \\ \hline \hline	\end{tabular}
	\label{10states}
\end{table}
For the NPSBR system, we propose an infinite Markov chain $X_k, k=0,1,\ldots$ of QBD type characterized as in \eqref{qbd} with state-space  $\{(i,j): \: 0 \leq i < \infty, \:1 \leq j \leq 10 \}$ with 10 phases. Table~\ref{10states} describes each of the ten phases used the NPSBR system.
The characterizing matrix $\bm{A_0}$ of the proposed QBD is given in Eqn.~\eqref{A0NPSBR}.
Moreover, $\bm{A_1}$ is matrix of zeros, $\bm{A_2}$
and $\bm{B_1}$ are matrices of zeros except for their $(10,10)^{\text{th}}$ entry which is one, and 
$\bm{B_0}$ is a matrix of zeros except for its $(10,1)^{\text{th}}$ entry which is $1-a$ and  $(10,2)^{\text{th}}$ entry which is $a$.

This QBD evolves in the form of repetitive cycles as in the previous two queueing sytems. A cycle begins with the arrival of a successful source-$1$ packet which then first waits in the waiting room (phase 1) which is then served (phase 2). Following phase 2, the QBD evolves until the completion of the next successful  source-$1$ packet's service completion (phases 3 to 9). Finally, in phase 10, we prepare for the next cycle. From the evolution of this QBD, we observe that the values that the level process of the proposed QBD takes in phases 5 to 9 in one QBD cycle, coincide with the sample values of the AoI process in its own cycles, as given in Table~\ref{illustrative}. 
We now present the following main result without proof for the NPSBR queueing system.
\begin{theorem}
	\label{thm:npb}
	Let $X_k \sim QBD(\bm{B_0},\bm{B_1},\bm{A_0},\bm{A_1},\bm{A_2})$ having the characterizing matrices as in \eqref{A0NPSBR} with its stationary vector for level $k$ of size $1 \times 10$ denoted by $\bm{\pi_k}$ being in matrix geometric form $
	\bm{\pi_k} = \bm{\pi_0} \bm{R}^k u_k$. Then, the steady-state pmf of the AoI sequence for the NPSBR system for source-$1$ is the same as that of $L^{\mathcal{S_A}}$ which is the steady-state level of the QBD restricted to the subset $\mathcal{S_A} = \{ 5,6,7,8,9\}.$ Moreover, the steady-state pmf of the PAoI sequence for the NPSBR system for source-$1$ is the same as that of $1+L^{\mathcal{S_P}}$ where  $L^{\mathcal{S_P}}$ is the steady-state level of the QBD restricted to phase 6, i.e., $\mathcal{S_P} = \{ 6 \}.$
\end{theorem}
%\begin{theorem}
%	\label{thm:npsbr}
%	Let $X_k \sim QBD(\bm{B_0},\bm{B_1},\bm{A_0},\bm{A_1},\bm{A_2})$ having the characterizing matrices as in \eqref{A0npb} with its stationary vector for level $k$ of size $1 \times 10$ denoted by $\bm{\pi_k}$ being in matrix geometric form $
%	\bm{\pi_k} = \bm{\pi_0} \bm{R}^k u_k$. Then, the steady-state pmfs of the AoI and PAoI sequences for the NPSBR system for source-$1$ are given with the following expression:
%	\begin{equation}
%	p_{\Delta}(\ell) = \bm{c_a} \bm{R}^{\ell} \bm{b_a} u_{\ell}, \
%	p_{\Phi}(\ell ) = \bm{c_p} \bm{R}^{\ell - 1} \bm{b_p} u_{\ell - 1},
%	\label{thmnpb}
%	\end{equation}
%	where
%	 \begin{align}
%	\bm{b_a} & =\begin{bmatrix} \bm{0_{1 \times 4}} &\bm{1_{1 \times 5}} & 0 \end{bmatrix} ^T, \ \bm{c_a}  = \bm{\pi_0}/ (\bm{\pi_0} (\bm{I} - \bm{R})^{-1} \bm{b_a}), \nonumber  \\ \bm{b_p} &=\begin{bmatrix}  \bm{0_{1 \times 5}} &
%		1 &  \bm{0_{1 \times 4}} \end{bmatrix} ^T,  \ \bm{c_p} = \bm{\pi_0}/ (\bm{\pi_0}(\bm{I} - \bm{R})^{-1} \bm{b_p}). \label{pmfnpsbr} 
%	\end{align}
%	Moreover, their expected values are given by:
%	\begin{align}
%	E[\Delta] & = \bm{c_a} \left( (\bm{I} - \bm{R})^{-2} - (\bm{I} - \bm{R})^{-1}) \right) \bm{b_a}, \nonumber \\
%	E[\Phi] & = \bm{c_p} \left( (\bm{I} - \bm{R})^{-2} - (\bm{I} - \bm{R})^{-1}) \right) \bm{b_p}. \label{meannpsbr}
%	\end{align}
%\end{theorem}

\section{Numerical Results}
\begin{figure}[htb]
	\centering
	\includegraphics[]{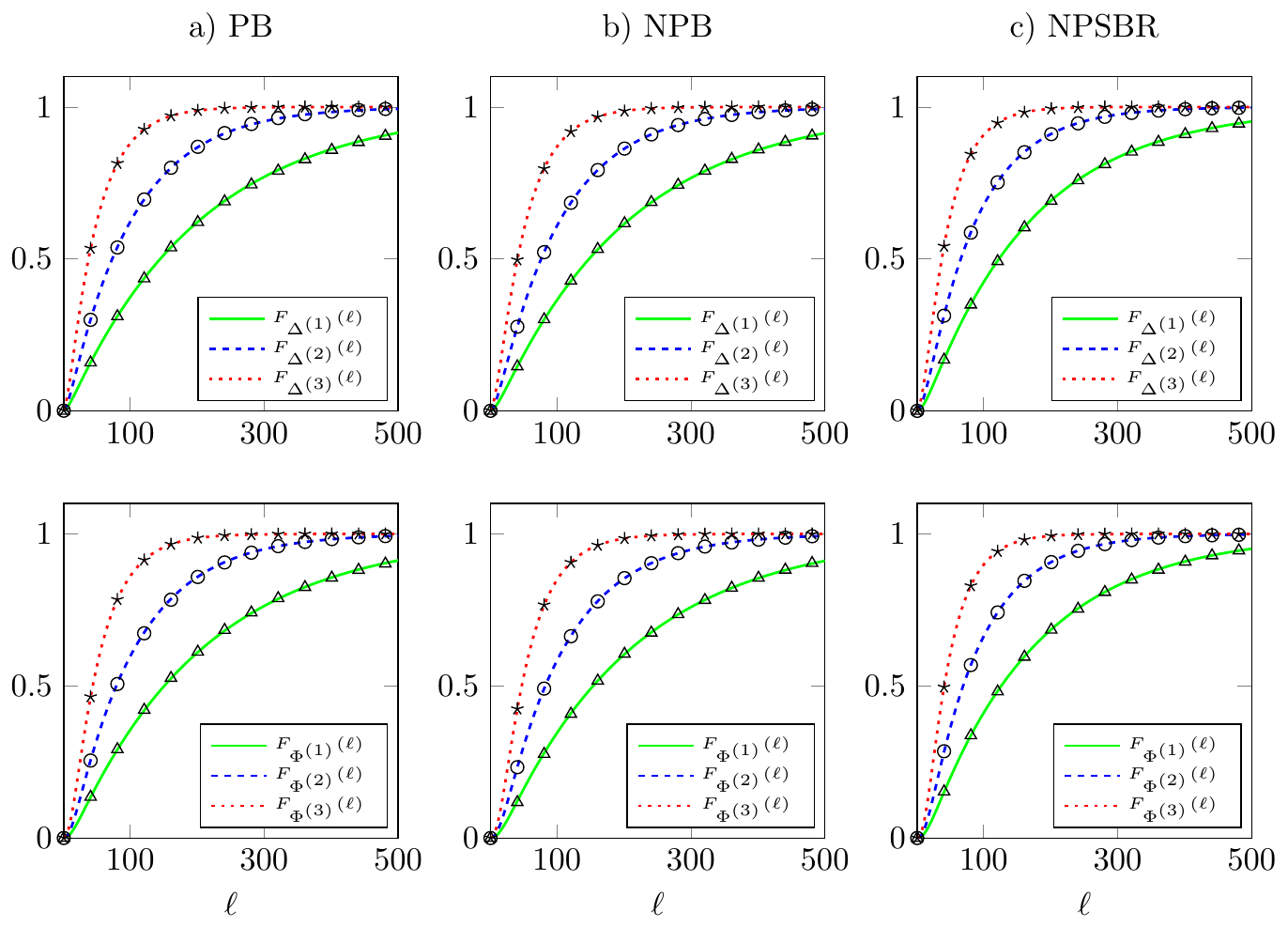}
	\caption{The cdfs of AoI and PAoI processes for three different queueing models obtained by the proposed models and simulations results (denoted by markers) when $q = 0.1$, $\rho = 0.5$, $p = \rho q$ and $(p_1,p_2,p_3)=(\frac{p}{7},\frac{2p}{7},\frac{4p}{7})$.}
	\label{fig1}
\end{figure}

\begin{figure}[thb]
	\centering
	\includegraphics[]{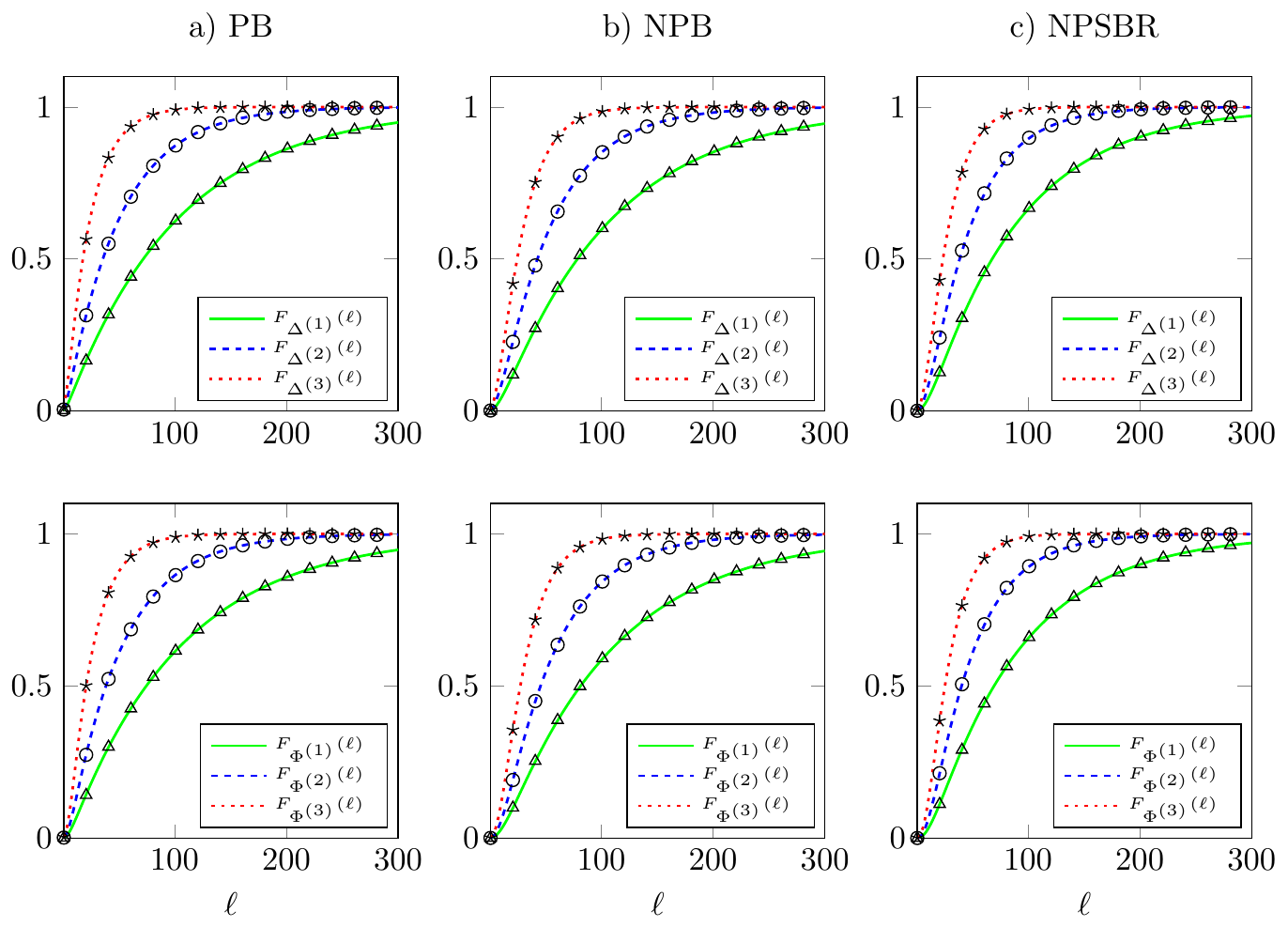}
	\caption{The cdfs of AoI and PAoI processes for three different queueing models obtained by the proposed models and simulations results(denoted by markers) when $q = 0.1$, $\rho = 2$,  $p = \rho q$ and $(p_1,p_2,p_3)=(\frac{p}{7},\frac{2p}{7},\frac{4p}{7})$. }
	\label{fig2}
\end{figure}

In the first numerical example, the analytical models we have proposed for the three queueing disciplines NPB, PB, and NPSBR, are validated by simulations. For this purpose, we
fix $N=3$, $q=0.1$, and for a given value of the load parameter $\rho$, we set the Bernoulli arrival vector of probabilities $(p_1,p_2,p_3)=(\frac{p}{7},\frac{2p}{7},\frac{4p}{7})$
where $p=q\rho$. The cdfs of the AoI and PAoI random sequences for the three different queueing schemes obtained with the proposed analytical model and simulations are depicted in figures~\ref{fig1} and \ref{fig2}, respectively, for $\rho=0.5$ and $\rho=2$, respectively. In all the cases, we have observed a perfect match between the results obtained with the proposed analytical model and the simulation results. 

\begin{align} 
	\bm{A_0} & = 
	\begin{pmatrix}
	\bar{b} & b & 0 & 0&0 & 0&0 & 0 & 0&0\\ 
	0 & \bar{q}\gamma_0 & \bar{q}\gamma_1 & \bar{q}\gamma_2& q\gamma_0& q\gamma_1& q\gamma_2& 0 & 0 &0\\ 
	0& 0& \bar{q}\gamma_{01} & \bar{q}\gamma_2 & 0 & q\gamma_{01} &q\gamma_2 &0 & 0& 0\\ 
	0& 0& \bar{q}\gamma_1& \bar{q}\gamma_{02} & 0& q\gamma_1& q\gamma_{02} & 0& 0 &0 \\ 
	0& 0& 0& 0& \gamma_0 & \gamma_1 & \gamma_2 & 0& 0&0 \\ 
	0& 0& 0& 0&0 & \bar{q} & 0& 0& 0& q \\ 
	0&0 & 0& 0& q\gamma_0& q\gamma_1& \bar{q}\gamma_{0} + q\gamma_2   & \bar{q}\gamma_1& \bar{q}\gamma_2 & 0\\ 
	0& 0& 0& 0& 0& q\gamma_{01}& q\gamma_2& \bar{q}\gamma_{01} & \bar{q}\gamma_2& 0\\ 
	0& 0& 0& 0& 0& q\gamma_1& q\gamma_{02} & \bar{q}\gamma_1& \bar{q}\gamma_{02} & 0\\ 
	0& 0& 0& 0& 0& 0& 0& 0& 0& 0\\ 
	\end{pmatrix},
	\label{A0NPSBR}
\end{align}

In the second numerical example, we study the impact of the choice of the per-source packet generation probabilities $p_1$ and $p_2$ using the analytical model only on the system cost $C(\alpha)$ for a two-source system, which is given in the following form:
\begin{equation}
C(\alpha)= E[\Delta^{(1)}] + \alpha E[\Delta^{(2)}], \ 0 \leq \alpha \leq 1, \label{cost}
\end{equation} 
which allows one to give more importance to source-$1$ over source-$2$ with a proper choice of the cost parameter $\alpha$. When $\alpha=1$, both sources are equally important whereas when $\alpha \rightarrow 0$, the mean age of the second source becomes less relevant. For a given cost parameter $\alpha$, we do exhaustive search to find the optimum packet generation probabilities $p_1^*$ and $p_2^*$ that minimize the system cost $C(\alpha)$ where the minimum attainable cost is denoted by $C^*(\alpha)$. Table~\ref{table1} tabulates the optimum values  $p_1^*$, $p_2^*$, $C^*(\alpha)$ for three different values of the service time parameter $q$ and for each of the three queueing disciplines of interest.
As a second scenario, for power budgeting, we impose a bound on the overall packet generation rate while doing exhaustive search, i.e., $p = p_1 + p_2 \leq \beta$, where $\beta$ is power constraint parameter since packet generation requires a certain energy consumption. In Table~\ref{table2}, we provide our results for the specific case of $\beta=0.1$. 
Studying tables~\ref{table1} and \ref{table2}, we have the following observations:
\begin{itemize}
	\item The optimum values of $p_1$ and $p_2$ turn out to be the same for all the three queueing disciplines but the optimum cost values are not necessarily the same for each discipline. 
	\item In the lack of a power constraint, the optimum packet generation rate for source-$1$ is always one and that of source-$2$ is one only when $\alpha=1$ and it decreases when $\alpha$ decreases. Since $p_1^*=1$ in all these cases, there is always a packet arrival at each time instant and the two queueing disciplines NPB and NPSBR behave exactly the same under the same set of traffic parameters. In this case, the PB always outperforms the other two disciplines when the optimum packet generation rates are employed thanks to the memoryless property of the geometric service time distribution.
	\item When there is a power constraint imposed on the packet generation rates, i.e., $\beta=0.1$, the situation is very different. In this case, NPSBR always outperforms NPB and there are situations when NPSBR outperforms PB which appear to occur for lower average service times.
\end{itemize} 

\section{Conclusions}
In this paper, we propose a discrete-time queueing model to derive the exact distributions of the AoI and PAoI sequences in a multi-source IoT-based status update system with Bernoulli packet generations and geometrically distributed service times. Three queueing disciplines are considered, namely 
non-preemptive bufferless, preemptive bufferless, and non-preemptive single buffer with replacement systems.
The proposed method gives rise to matrix geometric expressions for the related distributions and moreover, the factorial moments of AoI and PAoI of any order are explicitly given using matrix-vector operations.  
Numerical examples along with simulations are presented to validate the proposed approach.
Using the proposed analytical model, we also provide a numerical example on the optimum choice of the Bernoulli parameters in a practical IoT system with two sources with diverse AoI requirements and the three queueing disciplines are compared and contrasted in this setting. We have shown that the PB system yields the best performance in majority of the cases whereas it is slightly outperformed by the NPSBR discipline in scenarios with power constraints and relatively shorter average service times.

\renewcommand{\arraystretch}{1.2}
\begin{table}[]
	\centering
	\caption{The optimum values of $p_1$ and $p_2$ and the optimum value of $C(\alpha)$ for three different values of $q \in \{ 0.05,0.1,0.25 \}$ for various values of $\alpha$ for the three queueing disciplines in the absence of a power constraint.}
	\small\addtolength{\tabcolsep}{-2.5pt}
	\begin{tabular}{c|cc|c|cc|c|cc|c}
		\hline \hline
		\multicolumn{9}{c}{$q=0.05$} \\ \hline \hline
		&   \multicolumn{3}{c}{PB}& \multicolumn{3}{c}{NPB} & \multicolumn{3}{c}{NPSBR}  \\ \hline
		$\alpha$ & $p_1^*$ &  $p_2^*$ &  $C^*(\alpha)$ & $p_1^*$ &  $p_2^*$  &  $C^*(\alpha)$ & $p_1^*$ &  $p_2^*$    &  $C^*(\alpha)$   \\ \hline
		.1&$1$&$0.48$&$34.6$&$1$&$0.48$&$55.5$&$1$&$0.48$&$55.5$\\ \cline{1-10}
		.2&$1$&$0.62$&$41.9$&$1$&$0.62$&$64.7$&$1$&$0.62$&$64.7$\\ \cline{1-10}
		.3&$1$&$0.71$&$47.9$&$1$&$0.71$&$72.6$&$1$&$0.71$&$72.6$\\ \cline{1-10}
		.4&$1$&$0.77$&$53.3$&$1$&$0.77$&$79.9$&$1$&$0.77$&$79.9$\\ \cline{1-10}
		.5&$1$&$0.83$&$58.3$&$1$&$0.83$&$86.8$&$1$&$0.83$&$86.8$\\ \cline{1-10}
		.6&$1$&$0.87$&$63.0$&$1$&$0.87$&$93.4$&$1$&$0.87$&$93.4$\\ \cline{1-10}
		.7&$1$&$0.91$&$67.5$&$1$&$0.91$&$99.8$&$1$&$0.91$&$99.8$\\ \cline{1-10}
		.8&$1$&$0.94$&$71.8$&$1$&$0.94$&$106.0$&$1$&$0.94$&$106.0$\\ \cline{1-10}
		.9&$1$&$0.97$&$75.9$&$1$&$0.97$&$112.0$&$1$&$0.97$&$112.0$\\ \cline{1-10}
		1&$1$&$1.00$&$80.0$&$1$&$1.00$&$118.0$&$1$&$1.00$&$118.0$\\ \cline{1-10} \hline
		\hline
		\multicolumn{9}{c}{$q=0.1$} \\ \hline \hline
		&   \multicolumn{3}{c}{PB}& \multicolumn{3}{c}{NPB} & \multicolumn{3}{c}{NPSBR}  \\ \hline
		$\alpha$ & $p_1^*$ &  $p_2^*$ &  $C^*(\alpha)$ & $p_1^*$ &  $p_2^*$  &  $C^*(\alpha)$ & $p_1^*$ &  $p_2^*$    &  $C^*(\alpha)$   \\ \hline
		.1&$1$&$0.48$&$17.3$&$1$&$0.48$&$27.2$&$1$&$0.48$&$27.2$\\ \cline{1-10}
		.2&$1$&$0.62$&$20.9$&$1$&$0.62$&$31.7$&$1$&$0.62$&$31.7$\\ \cline{1-10}
		.3&$1$&$0.71$&$24.0$&$1$&$0.71$&$35.7$&$1$&$0.71$&$35.7$\\ \cline{1-10}
		.4&$1$&$0.77$&$26.6$&$1$&$0.77$&$39.2$&$1$&$0.77$&$39.2$\\ \cline{1-10}
		.5&$1$&$0.83$&$29.1$&$1$&$0.83$&$42.6$&$1$&$0.83$&$42.6$\\ \cline{1-10}
		.6&$1$&$0.87$&$31.5$&$1$&$0.87$&$45.9$&$1$&$0.87$&$45.9$\\ \cline{1-10}
		.7&$1$&$0.91$&$33.7$&$1$&$0.91$&$49.0$&$1$&$0.91$&$49.0$\\ \cline{1-10}
		.8&$1$&$0.94$&$35.9$&$1$&$0.94$&$52.1$&$1$&$0.94$&$52.1$\\ \cline{1-10}
		.9&$1$&$0.97$&$38.0$&$1$&$0.97$&$55.1$&$1$&$0.97$&$55.1$\\ \cline{1-10}
		1&$1$&$1.00$&$40.0$&$1$&$1.00$&$58.0$&$1$&$1.00$&$58.0$\\ \cline{1-10}
		\hline \hline
		\multicolumn{9}{c}{$q=0.25$} \\ \hline \hline
		&   \multicolumn{3}{c}{PB}& \multicolumn{3}{c}{NPB} & \multicolumn{3}{c}{NPSBR}  \\ \hline
		$\alpha$ & $p_1^*$ &  $p_2^*$ &  $C^*(\alpha)$ & $p_1^*$ &  $p_2^*$  &  $C^*(\alpha)$ & $p_1^*$ &  $p_2^*$    &  $C^*(\alpha)$   \\ \hline
		.1&$1$&$0.48$&$6.9$&$1$&$0.48$&$10.2$&$1$&$0.48$&$10.2$\\ \cline{1-10}
		.2&$1$&$0.62$&$8.4$&$1$&$0.62$&$12.0$&$1$&$0.62$&$12.0$\\ \cline{1-10}
		.3&$1$&$0.71$&$9.6$&$1$&$0.71$&$13.5$&$1$&$0.71$&$13.5$\\ \cline{1-10}
		.4&$1$&$0.77$&$10.7$&$1$&$0.77$&$14.9$&$1$&$0.77$&$14.9$\\ \cline{1-10}
		.5&$1$&$0.83$&$11.7$&$1$&$0.83$&$16.2$&$1$&$0.83$&$16.2$\\ \cline{1-10}
		.6&$1$&$0.87$&$12.6$&$1$&$0.87$&$17.4$&$1$&$0.87$&$17.4$\\ \cline{1-10}
		.7&$1$&$0.91$&$13.5$&$1$&$0.91$&$18.6$&$1$&$0.91$&$18.6$\\ \cline{1-10}
		.8&$1$&$0.94$&$14.4$&$1$&$0.94$&$19.8$&$1$&$0.94$&$19.8$\\ \cline{1-10}
		.9&$1$&$0.97$&$15.2$&$1$&$0.97$&$20.9$&$1$&$0.97$&$20.9$\\ \cline{1-10}
		1&$1$&$1.00$&$16.0$&$1$&$1.00$&$22.0$&$1$&$1.00$&$22.0$\\ \cline{1-10}
	\end{tabular}
	\label{table1}
\end{table}

%\ifCLASSOPTIONcaptionsoff
%\newpage
%\fi
%

\begin{table}[]
	\centering
	\caption{The optimum values of $p_1$ and $p_2$ and the optimum value of $C(\alpha)$ for three different values of $q \in \{ 0.05,0.1,0.25 \} $ for various values of $\alpha$ for the three queueing disciplines. The power constraint is taken as $p_1 + p_2 \leq \beta=0.1$.}
	\small\addtolength{\tabcolsep}{-2.5pt}
	\begin{tabular}{c|cc|c|cc|c|cc|c}
		\hline \hline
		\multicolumn{9}{c}{$q=0.05$} \\ \hline \hline
		&   \multicolumn{3}{c}{PB}& \multicolumn{3}{c}{NPB} & \multicolumn{3}{c}{NPSBR}  \\ \hline
		$\alpha$ & $p_1^*$ &  $p_2^*$ &  $C^*(\alpha)$ & $p_1^*$ &  $p_2^*$  &  $C^*(\alpha)$ & $p_1^*$ &  $p_2^*$    &  $C^*(\alpha)$   \\ 
		&  \multicolumn{2}{c|}{$(\times 10^{-2})$} &  & \multicolumn{2}{c|}{$(\times 10^{-2})$} &  & \multicolumn{2}{c|}{$(\times 10^{-2})$}     \\ \hline
		.1&$7.6$&$2.4$&$50.6$&$7.6$&$2.4$&$64.9$&$7.6$&$2.4$&$61.7$\\ \cline{1-10}
		.2&$6.9$&$3.1$&$61.2$&$6.9$&$3.1$&$76.8$&$6.9$&$3.1$&$72.0$\\ \cline{1-10}
		.3&$6.4$&$3.6$&$70.0$&$6.4$&$3.6$&$86.9$&$6.4$&$3.6$&$81.0$\\ \cline{1-10}
		.4&$6.1$&$3.9$&$77.9$&$6.1$&$3.9$&$96.1$&$6.1$&$3.9$&$89.2$\\ \cline{1-10}
		.5&$5.8$&$4.2$&$85.2$&$5.8$&$4.2$&$104.7$&$5.8$&$4.2$&$97.0$\\ \cline{1-10}
		.6&$5.6$&$4.4$&$92.1$&$5.6$&$4.4$&$112.9$&$5.6$&$4.4$&$104.4$\\ \cline{1-10}
		.7&$5.4$&$4.6$&$98.7$&$5.4$&$4.6$&$120.8$&$5.4$&$4.6$&$111.5$\\ \cline{1-10}
		.8&$5.3$&$4.7$&$105.0$&$5.3$&$4.7$&$128.4$&$5.3$&$4.7$&$118.5$\\ \cline{1-10}
		.9&$5.1$&$4.9$&$111.1$&$5.1$&$4.9$&$135.8$&$5.1$&$4.9$&$125.3$\\ \cline{1-10}
		1&$5.0$&$5.0$&$117.0$&$5.0$&$5.0$&$143.0$&$5.0$&$5.0$&$131.9$\\ \cline{1-10}
		\hline \hline
		\multicolumn{9}{c}{$q=0.1$} \\ \hline \hline
		&   \multicolumn{3}{c}{PB}& \multicolumn{3}{c}{NPB} & \multicolumn{3}{c}{NPSBR}  \\ \hline
		$\alpha$ & $p_1^*$ &  $p_2^*$ &  $C^*(\alpha)$ & $p_1^*$ &  $p_2^*$  &  $C^*(\alpha)$ & $p_1^*$ &  $p_2^*$    &  $C^*(\alpha)$   \\ 
		&  \multicolumn{2}{c|}{$(\times 10^{-2})$} &  & \multicolumn{2}{c|}{$(\times 10^{-2})$} &  & \multicolumn{2}{c|}{$(\times 10^{-2})$}     \\ \hline
		.1&$7.6$&$2.4$&$33.2$&$7.6$&$2.4$&$38.4$&$7.6$&$2.4$&$34.8$\\ \cline{1-10}
		.2&$6.9$&$3.1$&$40.3$&$6.9$&$3.1$&$45.9$&$6.9$&$3.1$&$41.0$\\ \cline{1-10}
		.3&$6.4$&$3.6$&$46.1$&$6.4$&$3.6$&$52.2$&$6.4$&$3.6$&$46.3$\\ \cline{1-10}
		.4&$6.1$&$3.9$&$51.3$&$6.1$&$3.9$&$57.8$&$6.1$&$3.9$&$51.2$\\ \cline{1-10}
		.5&$5.8$&$4.2$&$56.1$&$5.8$&$4.2$&$63.1$&$5.8$&$4.2$&$55.7$\\ \cline{1-10}
		.6&$5.6$&$4.4$&$60.6$&$5.6$&$4.4$&$68.1$&$5.6$&$4.4$&$60.1$\\ \cline{1-10}
		.7&$5.4$&$4.6$&$65.0$&$5.4$&$4.6$&$72.9$&$5.4$&$4.6$&$64.2$\\ \cline{1-10}
		.8&$5.3$&$4.7$&$69.1$&$5.3$&$4.7$&$77.5$&$5.3$&$4.7$&$68.3$\\ \cline{1-10}
		.9&$5.1$&$4.9$&$73.1$&$5.1$&$4.9$&$82.0$&$5.1$&$4.9$&$72.2$\\ \cline{1-10}
		1&$5.0$&$5.0$&$77.0$&$5.0$&$5.0$&$86.4$&$5.0$&$5.0$&$76.0$\\ \cline{1-10}
		\hline \hline
		\multicolumn{9}{c}{$q=0.25$} \\ \hline \hline
		&   \multicolumn{3}{c}{PB}& \multicolumn{3}{c}{NPB} & \multicolumn{3}{c}{NPSBR}  \\ \hline
		$\alpha$ & $p_1^*$ &  $p_2^*$ &  $C^*(\alpha)$ & $p_1^*$ &  $p_2^*$  &  $C^*(\alpha)$ & $p_1^*$ &  $p_2^*$    &  $C^*(\alpha)$   \\ 
		&  \multicolumn{2}{c|}{$(\times 10^{-2})$} &  & \multicolumn{2}{c|}{$(\times 10^{-2})$} &  & \multicolumn{2}{c|}{$(\times 10^{-2})$}     \\ \hline
		.1&$7.6$&$2.4$&$22.8$&$7.6$&$2.4$&$23.8$&$7.6$&$2.4$&$21.9$\\ \cline{1-10}
		.2&$6.9$&$3.1$&$27.7$&$6.9$&$3.1$&$28.8$&$6.9$&$3.1$&$26.3$\\ \cline{1-10}
		.3&$6.5$&$3.5$&$31.7$&$6.5$&$3.5$&$32.9$&$6.5$&$3.5$&$29.9$\\ \cline{1-10}
		.4&$6.1$&$3.9$&$35.3$&$6.1$&$3.9$&$36.6$&$6.1$&$3.9$&$33.2$\\ \cline{1-10}
		.5&$5.9$&$4.1$&$38.6$&$5.9$&$4.1$&$40.0$&$5.9$&$4.1$&$36.2$\\ \cline{1-10}
		.6&$5.6$&$4.4$&$41.7$&$5.6$&$4.4$&$43.2$&$5.6$&$4.4$&$39.1$\\ \cline{1-10}
		.7&$5.4$&$4.6$&$44.7$&$5.4$&$4.6$&$46.3$&$5.4$&$4.6$&$41.8$\\ \cline{1-10}
		.8&$5.3$&$4.7$&$47.6$&$5.3$&$4.7$&$49.2$&$5.3$&$4.7$&$44.5$\\ \cline{1-10}
		.9&$5.1$&$4.9$&$50.3$&$5.1$&$4.9$&$52.1$&$5.1$&$4.9$&$47.1$\\ \cline{1-10}
		1&$5.0$&$5.0$&$53.0$&$5.0$&$5.0$&$54.8$&$5.0$&$5.0$&$49.6$\\ \cline{1-10}
	\end{tabular}
	\label{table2}
\end{table}

\bibliographystyle{ieeetran}
\bibliography{AoIDiscrete}

\end{document}